\documentclass[twocolumn,prl,showpacs]{revtex4}
\usepackage{graphicx,color}
\begin{document}

\title{
Superconductivity in SrFe$_{2-x}$Co$_x$As$_2$:
Internal Doping of the Iron Arsenide Layers
}

\author{A.\ Leithe-Jasper}
\author{W.\ Schnelle}
\author{C.\ Geibel}
\author{H.\ Rosner}

\affiliation{Max-Planck-Institut f\"ur Chemische Physik fester Stoffe,
N\"othnitzer Str.\ 40, 01187 Dresden, Germany}

\begin{abstract}
In the electron doped compounds SrFe$_{2-x}$Co$_x$As$_2$
superconductivity with $T_\mathrm{c}$ up to 20\,K is observed for
$0.2\leq x \leq 0.4$. Results of structure determination, magnetic
susceptibility, electrical resistivity, and specific heat are
reported. The observation of bulk superconductivity in all
thermodynamic properties -- despite strong disorder in the Fe-As
layer -- favors an itinerant picture in contrast to the cuprates and
renders a $p$- or $d$-wave scenario unlikely. DFT calculations find
that the substitution of Fe by Co ($x\geq 0.3$) leads to the
suppression of the magnetic ordering present in SrFe$_2$As$_2$ due to
a rigid down-shift of the Fe-3$d_{x^2-y^2}$ related band edge in the
density of states.
\end{abstract}


\maketitle

The observation of superconductivity in layered Fe-As systems has
recently attracted considerable interest. The undoped compounds,
e.g., LaFeAsO \cite{Kamihara08a,delaCruz08aetal} or $A$Fe$_2$As$_2$
($A$ = Ba, Sr) \cite{Rotter08a,Krellner08a,Tegel08a}, present an
antiferromagnetic (afm) transition related to a structural
transition at temperatures in the range 140\,K to 205\,K. Upon
doping both the structural and the magnetic ordering get suppressed
and superconductivity with $T_\mathrm{c}$ up to 56\,K appears
\cite{Kamihara08a,CWang08betal,Rotter08b}. The onset of
superconductivity at the disappearance of an afm ordered state
induced by charge doping from a reservoir layer is reminiscent of
the behavior observed in the cuprates. Since furthermore Fe and Cu
are both 3$d$ elements, and because of similarities in the structure
of the layers, an analogy between the high temperature
superconductivity (HTSC) in the cuprates and the layered Fe-As
systems was suggested in a large number of reports. On the other
hand, the absence of a Mott-insulator transition as well as the
small ordered Fe moment in the undoped compounds put this analogy
into question, and suggest that a model based on itinerant and
weakly correlated 3$d$ electrons might be more appropriate.

One way to get deeper insight into these questions is a doping
experiment on the Fe site. Presently, doping was generally performed
on sites in-between the Fe-As layers, either on the $R$-site or the
O-site in the $R$FeAsO systems or on the $A$ site in the
$A$Fe$_2$As$_2$ compounds. Such kind of doping corresponds, both in
a localized and in an itinerant model, to a simple charge doping and
is therefore not very suitable for discriminating between both
models. On the other hand, replacing a small amount of Fe by another
3$d$ element should lead to different effects depending on the model
used. In an itinerant model, within a simple rigid-band approach,
substitution on the Fe site is expected to be quite similar to
indirect doping via the interlayer sites, since only the total
number of electrons is relevant. In a picture with localized 3$d$
electrons, doping on the Fe site should lead to a very different
behavior, since the correlations in the 3$d$ layers are directly
affected -- a few percent Ni or Zn doping on the Cu site in a HTSC
cuprate lead to a drastic reduction of $T_\mathrm{c}$.

Therefore, we investigated thermodynamic properties and electrical
conductivity of the solid solution SrFe$_{2-x}$Co$_x$As$_2$. While
pure SrFe$_2$As$_2$ undergoes a lattice distortion and afm ordering
at $T_0$ = 205\,K \cite{Krellner08a}, a small amount of Co
substitution leads to a rapid decrease of $T_0$, followed by the
onset of superconductivity in the concentration range $0.2 \leq x
\leq 0.4$ with a maximum $T_\mathrm{c}$ of $\approx 20$\,K. The
observation of superconductivity in Co substituted SrFe$_2$As$_2$
provides strong evidence that an analogy with the HTSC cuprates is
not really justified. We therefore performed band structure
calculations and discuss the electronic structure in view of our
experimental results. The good agreement between calculations and
experimental observations confirms that an itinerant approach is
appropriate for these layered Fe-As systems. The observation of
superconductivity despite a strong in-layer disorder puts also a
strong constraint on the possible superconducting order parameters.
While finalizing our investigation, Sefat \textit{et
al.}\,\cite{Sefat08aetal} and Wang \textit{et
al.}\,\cite{CWang08aetal} reported the observation of
superconductivity in Co doped LaFeAsO with a maximum $T_\mathrm{c}
\approx 10$\,K. While their results support our analysis, the higher
$T_\mathrm{c}$ observed in SrFe$_{2-x}$Co$_x$As$_2$ establishes much
stronger evidence.

Samples were prepared by a sintering technique in glassy-carbon
crucibles which were welded into tantalum containers and sealed into
evacuated quartz tubes for heat treatment at 900\,$^\circ$C for
16\,h followed by two regrinding and compaction steps. First
precursors SrAs, Co$_2$As and Fe$_2$As were synthesized from
elemental powders sintered at 600\,$^\circ$C for 48\,h (Fe, Co
powder, 99.9 wt.{\%}; As sublimed lumps 99.999 wt.{\%}; Sr 99.99
wt.{\%}. These educts were then powdered, blended in stoichiometric
ratios, compacted, and heat treated. All steps were carried out
within an argon-filled glove box (O$_2$, H$_2$O $< 1$\,ppm). This
mode of synthesis helps to prevent possible contamination by toxic
arsenic \cite{Nriagu94}. Samples were obtained in the form of
sintered pellets.

Metallographic preparations and microstructure investigation on
polished surfaces with optical microscopy were performed in an
argon-filled glove box. Electron-probe microanalysis (EPMA) with
energy dispersive analysis was accomplished in a Philips XL30
scanning electron microscope. Powder X-ray diffraction was performed
using Co K$\alpha$1 radiation ($\lambda$ = 1.789007\,\AA) applying
the Guinier Huber technique with LaB$_6$ as internal standard ($a$ =
4.15692\,\AA). Crystallographic calculations were done with the
WinCSD program package \cite{WinCSD}.

Evaluation of the lattice parameters and EPMA investigations of
selected samples unambiguously reveals the substitution of Fe by Co
which is accompanied by a decrease of the $c$ axis length of the
unit cell (see Table \ref{thetable}). For the sample
SrFe$_{1.8}$Co$_{0.2}$As$_2$ which was essentially single phase
(minor impurity SrAs) the crystal structure was refined by Rietfeld
methods (BaAl$_4$-type of structure \cite{Andress35}, space group
$I4/mmm$, Sr in $2a$ (0, 0, 0), Fe/Co in $4d$ (0, 1/2, 1/4), As in
$4e$ (0, 0, 0.3613(1)); $R_I$ = 7.4, $R_P$ = 10.1). Other samples of
this investigation contained as minor impurity phases
Fe$_{1-x}$Co$_x$As.

Magnetization was measured from 1.8--400\,K in a SQUID magnetometer
in various fields both after zero-field cooling (zfc) and during
cooling in field (fc). Heat capacity was determined by a
relaxation-type method on a commercial measurement system
(1.9--100\,K, 0--9\,T). Electrical dc resistivity data $\rho(T)$
(3.8-320\,K) were collected with a current density of $< 1 \times
10^{-3}$\,A\,mm$^{-2}$ in a four contact arrangement. Due to the
sample geometry the inaccuracy in $\rho$ is estimated to be
$\pm$20\,{\%}.

To investigate the influence of electron doping on the electronic
structure of SrFe$_2$As$_2$ on a microscopic level, we performed
density functional band structure calculations within the local
(spin) density approximation (L(S)DA). Using the experimental
structural parameters \cite{Krellner08a}, we applied the
full-potential local-orbital code FPLO \cite{Koepernik99} (version
7.00-28) with the Perdew-Wang exchange correlation potential
\cite{PerdewWang92} and a carefully converged $k$-mesh of 24$^3$
points for the Brillouin zone. The substitution of Fe by Co and Sr
by La was simulated within the virtual crystal approximation (VCA).

\begin{figure}[htb]
\includegraphics[height=3.4in,angle=90]{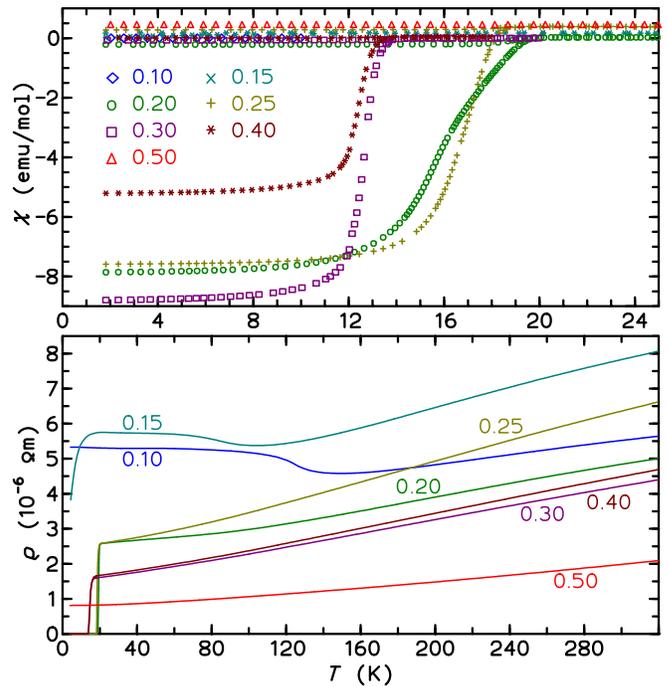}
\caption{(Color online)
Top: magnetic susceptibility $\chi(T)$ of SrFe$_{2-x}$Co$_x$As$_2$
samples in a nominal field of $\mu_0H$ = 2\,mT. Bottom: electrical
resistivity of the same samples.
\label{figchirho}}
\end{figure}

\begin{table}[htb]
\begin{center}
\caption{Lattice parameters $a,c$ of SrFe$_{2-x}$Co$_x$As$_2$
(nominal compositions) and superconducting transition temperature
$T_\mathrm{c}^\mathrm{mag}$ (defined as the crossing of the tangent
to the steepest slope of the fc transition with $\chi$ = 0),
$\chi_0$ is the high-field susceptibility at $T$=0. Data for $x$=0
are from Ref.\ \cite{Krellner08a}.
\label{thetable}}
\begin{ruledtabular}
\begin{tabular}{ccccc}
$x$  & $a$        & $c$        &$T_\mathrm{c}^\mathrm{mag}$& $\chi_0$           \\
     & (\AA)      & (\AA)      & (K)                       & (10$^{-6}$emu/mol) \\ \hline
0.00 & 3.924(3)   & 12.34(1)   & --                        & --                 \\
0.10 & 3.9291(1)  & 12.3321(7) & --                        & n.a.               \\
0.15 & 3.9272(1)  & 12.3123(5) & $<$1.8                    & $+$780             \\
0.20 & 3.9278(2)  & 12.3026(2) & 19.2                      & $+$600             \\
0.25 & 3.9296(2)  & 12.2925(9) & 18.1                      & ($+$850)           \\
0.30 & 3.9291(2)  & 12.2704(8) & 13.2                      & $+$460             \\
0.40 & 3.9293(1)  & 12.2711(7) & 12.9                      & $+$360             \\
0.50 & 3.9287(2)  & 12.2187(9) & $<$1.8                    & n.a.               \\
1.00 & 3.9618(1)  & 11.6378(6) & --                        & n.a.               \\
\end{tabular}
\end{ruledtabular}
\end{center}
\end{table}
\noindent

The magnetic susceptibility in a nominal field $\mu_0H$ = 2\,mT
(Fig.\ \ref{figchirho} top) displays strong diamagnetic signals in
zfc due to superconducting transitions in the $x$=0.2--0.4 samples
with onset temperatures $T_\mathrm{c}^\mathrm{mag}$ up to 19.2\,K
for $x$=0.2 (see Table \ref{thetable}). The compounds with $x$=0.1
and $x$=0.5 show no traces of superconductivity. Even for this low
external field the transitions are slightly rounded which indicates
that the samples are not fully homogeneous. While the shielding
signal (zfc) corresponds to the whole sample volume (considering
approximate demagnetizing factors), the Meissner effect (fc) is much
smaller, which is probably due to strong pinning.

The normal-state susceptibilities of the samples $x$=0.2, 0.3, and
0.4 are field-independent for $T > 200$\,K. There, the compounds
display a paramagnetism which can be described by $\chi(T) = \chi_0
+ \chi_2T^2$. The extrapolated $\chi_0$ values are given in
Table\,\ref{thetable}. Diamagnetic core contributions are
comparatively small ($\approx -50 \times 10^{-6}$ emu/mol). Due to
the current sample quality some caution is necessary in interpreting
$\chi_0$, but the order of magnitude and the temperature dependence
is typical for an enhanced Pauli paramagnet. Comparing the $\chi_0$
values with the electronic density of states at the Fermi level
obtained from band structure calculations (see below) suggests an
enhanced Sommerfeld-Wilson ratio, which decreases with increasing Co
content for $x\geq$0.15, in accordance with a magnetic instability
for $x\leq$0.15. SrCo$_2$As$_2$ is a Curie-Weiss paramagnet
($\mu_\mathrm{eff}$ = 2.06\,$\mu_\mathrm{B}$/f.u.,
$\theta_\mathrm{CW}$ = $-$29\,K) and does not show magnetic ordering
above 1.8\,K.

The temperature dependence of the resistivity $\rho(T)$
(Fig.\ \ref{figchirho} bottom) at high temperatures is generally that
of a bad metal. $\rho$(300\,K) decreases gradually with electron
doping from 5.5--8.0\,$\mu\Omega$\,m for $x$=0.10--0.15 to
2.0\,$\mu\Omega$\,m for $x$=0.50, as also observed for K-substituted
samples.\cite{Rotter08a} For $x$=0.10 a stepwise increase of $\rho(T)$
is observed below $\approx$130\,K which shifts to $\approx$90\,K for
$x$=0.15. This anomaly can be assigned to the afm ordering and
the related lattice distortion observed at $T_0$=205\,K in SrFe$_2$As$_2$
\cite{Krellner08a,Tegel08a}. Thus, Co substitution (electron doping)
leads to the suppression of the afm order in a similar way as reported
for the K-substitution (hole doping) \cite{GFChen08aetal}. For
$x$=0.20 no such anomaly is seen and the compound shows a
superconducting transition at 19.4\,K.

\begin{figure}[htb]
\includegraphics[width=3.4in,angle=0]{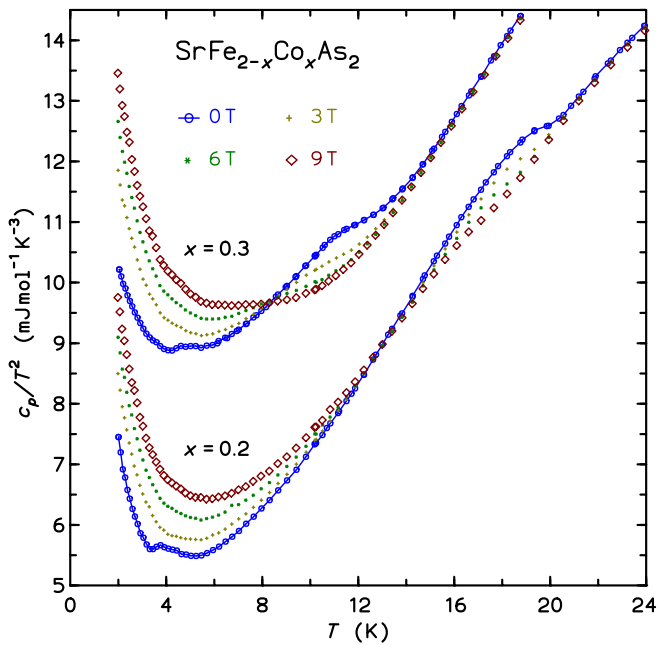}
\caption{(Color online) Molar isobaric specific heat $c_p/T^2$ of
SrFe$_{2-x}$Co$_x$As$_2$ samples ($x$=0.2; $x$=0.3 shifted up
by 3 units) for different magnetic fields. For better distinction the
data for $\mu_0H = 0$ are connected by a line.\label{figcp}}
\end{figure}

In Fig.\ \ref{figcp} the specific heat $c_p(T)$ for two selected
samples is presented in a $c_p(T)/T^2$ versus $T$ plot. Strongly
rounded anomalies with onset at $T_\mathrm{c}$ as determined by
$\rho(T)$ and $\chi(T)$ data are clearly visible confirming that the
superconductivity is a bulk phenomen in these samples. The idealized
step heights $\Delta c_p$ and the transition temperatures
$T_{c}^\mathrm{cal}$ were evaluated by a fit with a phononic
background and an electronic contribution according to the BCS
theory or the phenomenological two-liquid model which applies well
to the thermodynamic properties of some strong coupling
superconductors. The model is convoluted with a Gaussian to simulate
the broadening due to the chemical inhomogeneities.

$\Delta c_p/T_\mathrm{c}^\mathrm{cal}$ is $\approx$10 mJ mol$^{-1}$
K$^{-2}$ for $x$=0.2 and $\approx$13 mJ mol$^{-1}$ K$^{-2}$ for
$x$=0.3. Interestingly, well below $T_\mathrm{c}$ the specific heat
can still be described with a linear plus a $T^3$ Debye lattice
term, i.e.\ $c_p(T) = \gamma' T + \beta T^3$. While $\beta$ is
field-independent and corresponds to a lattice term with initial
Debye temperature $\Theta_\mathrm{D}(0)$ = 255(2)\,K for both
$x$=0.2 and 0.3, the linear term $\gamma'$ increases nearly linearly
with field. For $x$=0.2 the values of $\gamma'$ range from 12.6 mJ
mol$^{-1}$ at $\mu_0H$ = 0\,T to 17.5 at 9\,T. Whether the residual
$\gamma'$ is induced by defects (see e.g.
Ref.\,\cite{TrisconeJunod96}) or whether it is an intrinsic
contribution due to ungapped parts of the Fermi surface
\cite{Drechsler03aetal} has to be elucidated by further experiments.
While we cannot make a definite statement on the superconducting
coupling strength the better fit is obtained for $x$=0.2 and the
two-fluid model while for $x$=0.3 the BCS fit is superior. This
might indicate a variation of the coupling strength with Co content.

\begin{figure}[htb]
\includegraphics[width=3.4in,angle=0]{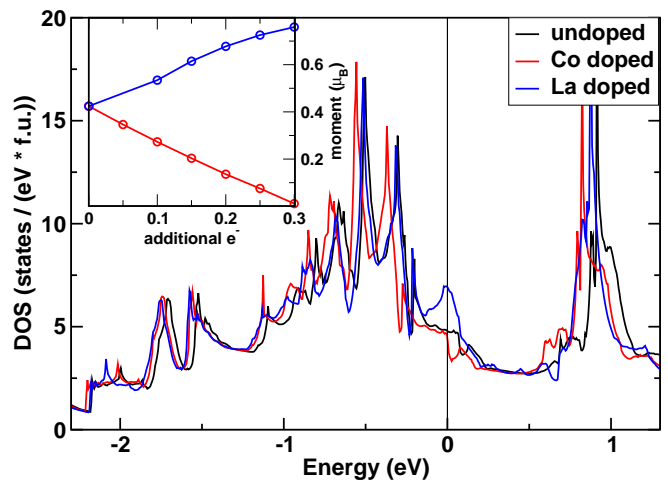}
\caption{(Color online)
Total electronic densities of states for SrFe$_2$As$_2$ (black),
SrFe$_{1.7}$Co$_{0.3}$As$_2$ (red) and the fictitious
Sr$_{0.7}$La$_{0.3}$Fe$_2$As$_2$ (blue) in the vicinity of the Fermi
level. Inset: dependence of the magnetic moment on the number of
additionally doped electrons per formula unit for doping at the Fe
site (red) and at the Sr site (blue), respectively.\label{dosfig}}
\end{figure}

For a microscopic study of the influence of doping with electrons on
the electronic structure in the vicinity of the Fermi level
$\varepsilon_\mathrm{F}$, we simulated two different kinds of
partial constituent exchange: (i) Co substitution at the Fe site and
(ii) a fictitious La substitution at the Sr site to study possible
differences for doping (i) within and (ii) outside the FeAs layers.
Throughout all calculations we used the experimental lattice
parameters \cite{Krellner08a} of the undoped SrFe$_2$As$_2$ to
separate the influence of geometry changes and pure electronic
doping effects. Since the changes of the lattice parameters upon
doping are very small (see Table\,\ref{thetable}) this approach is
well justified. The resulting total density of states (DOS) for the
Fe-3$d$ dominated regions of the valence band for
SrFe$_{2-x}$Co$_{x}$As$_2$ and Sr$_{1-x}$La$_{x}$Fe$_2$As$_2$,
respectively, is shown in Fig.\ \ref{dosfig} for an exemplary doping
level of $x$=0.3.

The two kinds of doping result in a rather different behavior.
Whereas (i) Co doping on the iron site results in an almost
perfectly rigid shift of the DOS for all studied doping levels
between $x$=0 and $x$=0.5, (ii) exchange of Sr by La can't be
described at all in a rigid band picture in the vicinity of
$\varepsilon_\mathrm{F}$ (see Fig.\ \ref{dosfig}). In the case of Co
substitution, a detailed analysis of the orbital-resolved DOS, the
corresponding bands and band characters shows that the rigid band
picture holds even for the individual orbitals close to
$\varepsilon_F$. In contrast, the substitution (ii) of Sr by La
shifts down the energy and changes the dispersion of a band with
Fe-As-Sr hybrid character in a way that it gets partially occupied,
whereas other bands remain basically unchanged. In consequence, this
leads to the formation of a quite pronounced peak at
$\varepsilon_\mathrm{F}$ and a sizable increase of the DOS (see
Fig.\ \ref{dosfig}). In general, from the experimental results for
the $R$FeAsO and the $A$Fe$_2$As$_2$ compound family it seems that
the appearance of superconductivity at low temperatures is
intrinsically related to the prior destruction of the spin density
wave. Therefore, we studied the instability towards magnetism
depending on the doping level for both doping scenarios (i) and
(ii). In Fig.\ \ref{dosfig} we plot the size of the ordered Fe
moment corresponding to the minimum in total energy in the VCA-LSDA
calculations as a function of Co or La substitution. (ii) While La
substitution stabilizes the magnetic state, electron doping by (i)
Co leads to the disappearance of the magnetic ground state at
$x$=0.3.

The disappearance of the magnetic instability is intimately
connected with the occupation of the Fe-3$d_{x^2-y^2}$ related band
along the $\Gamma$-Z direction as previously demonstrated
\cite{Krellner08a} for the undoped compound (depending on the As $z$
position). For $x$=0.3 (see Fig.\ \ref{dosfig}) the related band
edge is situated right at $\varepsilon_\mathrm{F}$, leading to a
reduced DOS and thus to the destruction of the magnetic state. In
contrast, the tendency towards magnetism would be increased for (ii)
La substitution. The DOS enhancement upon La substitution might lead
to an instability of the Sr$_{1-x}$La$_x$Fe$_2$As$_2$ phase, which
could be the reason why such compositions have not yet been
reported.

The non-magnetic ground state for $x \leq 0.3$ in
SrFe$_{2-x}$Co$_{x}$As$_2$ in the VCA-LSDA calculations is in
surprisingly good agreement with the experimentally observed
suppression of the magnetic state and the onset of superconductivity
for $x$=0.2. In fact, the used VCA should result in an overestimate
of the stability of a magnetic ground state since it does not
include the direct influence of disorder in the Fe-As layer. Most
likely, the doped Co impurities are in a non-magnetic state,
indicated by calculations for SrCo$_2$As$_2$ that yield a non
magnetic ground state in agreement with our measurements. Thus, a
more sophisticated treatment of the Co impurities should lead to a
sizable reduction of the critical concentration $x$ for the
suppression of the magnetic instability, in line with our
experimental results.

In summary, we synthesized polycrystalline SrFe$_{2-x}$Co$_x$As$_2$
samples and investigated how electron doping within the Fe-As layers
affects the magnetic and electronic properties of this kind of
compounds. We found that Co substitution suppresses the afm
transition and the associated lattice distortion quite rapidly, from
$T_0$ = 205\,K at $x$=0 to $T_0 \approx 90$\,K at $x$=0.15. For
$x$=0.2 afm order is destroyed and bulk superconductivity with a
$T_\mathrm{c} \approx 20$\,K appears. Further increase of the Co
content leads to a reduction of $T_\mathrm{c}$ and finally to the
disappearance of superconductivity for $x \leq 0.5$. Thus electron
doping within the Fe-As layer by substituting Co for Fe has a
similar effect as hole doping in-between the layers by substituting
K for Sr, or as electron doping by substituting F for O in LaFeAsO.

The observation of superconductivity upon substitution on the 3$d$
site in the layered Fe-As systems is in strong contrast to the
behavior in cuprates, where substitution on the Cu site lead to a
fast suppression of superconductivity. This indicates that the
suggested analogy between HTSC in cuprates and in Fe-As systems is
not appropriate. Instead we observe a rigid-band like shift of the
DOS to lower energy upon Co substitution. The suppression of the afm
state, which occurs in the calculations at $x$=0.3, is connected
with the $d_{x^2-y^2}$ states being pushed below the Fermi level. In
contrast electron doping by substituting La for Sr leads in the
calculation to a change in the dispersion and energy of a Fe-As-Sr
hybrid band. The resulting increase of the DOS at
$\varepsilon_\mathrm{F}$ stabilizes the afm state and might induce
an instability of the SrFe$_2$As$_2$ phase. The nice correspondence
between the experiments and VCA-LSDA band structure calculations
indicates that an approach based on itinerant, weakly correlated
3$d$ electrons is more appropriate for the description of the Fe-As
systems than an approach based on localized 3$d$ moments within a
Hubbard model.

\begin{acknowledgments}
We thank U.\ Burkhardt, T.\ Vogel, R.\ Koban, K.\ Kreutziger, Yu.\
Prots, and R.\ Gumeniuk for assistance.
\end{acknowledgments}

\end{document}